\documentclass[aps,pre,twocolumn,showpacs,superscriptaddress,groupedaddress]{revtex4-1}
\pdfoutput=1
\usepackage{hyperref}
\usepackage{epsfig}
\usepackage{amsmath}
\usepackage{graphicx}
\usepackage{wrapfig}
\usepackage{dcolumn}
\usepackage{bm}
\usepackage{amssymb}
\usepackage{titlesec}
\usepackage{mathtools}
\usepackage{relsize}
\usepackage{cleveref}
\usepackage{calligra}
\usepackage[usenames,dvipsnames]{color}

\begin{document}
\title{Extreme outbreak dynamics in epidemic models}
\author{Jason Hindes$^{1}$, Michael Assaf$^{\;2,3}$, and Ira B. Schwartz$^{1}$}
\affiliation{$^{1}$U.S. Naval Research Laboratory, Washington, DC 20375, USA}
\affiliation{$^{2}$Racah Institute of Physics, Hebrew University of Jerusalem, Jerusalem 91904, Israel}
\affiliation{$^{3}$Institute for Physics and Astronomy, University of Potsdam, Potsdam 14476, Germany}

\begin{abstract}
\textcolor{black}{Motivated by recent epidemic outbreaks, including those of COVID-19, we solve the canonical problem of calculating the dynamics and likelihood of extensive outbreaks in a population within a large class of stochastic epidemic models with demographic noise, including the Susceptible-Infected-Recovered (SIR) model and its general extensions.} In the limit of large populations, we compute the probability distribution for all extensive outbreaks, including those that entail unusually large or small (extreme) proportions of the population infected.
%Due to demographic noise, our analysis reveals that each extensive outbreak implies a unique, depletion or boost in the pool of susceptibles and an increase or decrease in the effective recovery rate compared to the usual mean-field dynamics.
\textcolor{black}{Our approach reveals that, unlike other well-known examples of rare events occurring in discrete-state stochastic systems, the statistics of extreme outbreaks emanate from a full continuum of Hamiltonian paths, each satisfying unique boundary conditions with a conserved probability flux.}
\end{abstract}

\maketitle

%\section{\label{sec:Intro}Introduction}
%The total number of infected individuals is a basic measure of the severity of an outbreak.
\textit{Introduction.} \textcolor{black}{Epidemic models are useful for understanding the general dynamics of infectious diseases, rumors, election outcomes, fads, and computer viruses\cite{keeling:infectious_diseases,AnderssonBook,RevModPhys.87.925,rodrigues2016application,billings2002unified,RevModPhys.80.1275,hindes2019degree,10.1137/19M1306658}.
Moreover, in the early days of emerging disease outbreaks, such as the current COVID-19 pandemic, societies rely on epidemics models for disease forecasting, as well as identifying the most effective control strategies\cite{ModelingCOVID-19,Ray2020.08.19.20177493,PhysRevE.103.L030301,Catching2020.08.12.20173047}}. To this end it is useful to quantify the risks of local epidemic outbreaks of various sizes. Within a given population, outbreak dynamics are typically described in terms of compartmental models\cite{keeling:infectious_diseases,MathematicalepidemiologyPastPresenFuture,rodrigues2016application}. For example, starting from some seed infection, over time individuals in a population make transitions between some number of
discrete disease states (susceptible, exposed, infectious, etc.) based on prescribed probabilities for a particular
disease\cite{ModelingCOVID-19,Ray2020.08.19.20177493,Catching2020.08.12.20173047,doi:10.1146/annurev-statistics-061120-034438,SIRSi,miller2019distribution}.
%The latter are fit from, e.g., incidence curves, positive test rates, etc\cite{ModelingCOVID-19,Ray2020.08.19.20177493,Catching2020.08.12.20173047,doi:10.1146/annurev-statistics-061120-034438,SIRSi,miller2019distribution}.
In the limit of infinite populations the
stochastic dynamics approach deterministic (mean-field) differential equations for the expected fraction of a population in each state\cite{keeling:infectious_diseases,MathematicalepidemiologyPastPresenFuture,rodrigues2016application,StochasticEpidemicModels}.
%Namely, given an initial condition, the dynamics of outbreaks that result in
%a finite fraction of the population infected are uniquely determined by solving a mean-field system.

Yet for real finite populations, outbreak dynamics have a wide range of different outcomes
for each initial condition, which are not predicted by mean-field models. \textcolor{black}{A natural and canonical question (for both statistical physics and population dynamics) is, what is the distribution of outbreak sizes?} Beside stochastic simulations\cite{doi:10.1146/annurev-statistics-061120-034438,keeling:infectious_diseases,StochasticEpidemicModels,doi:10.1098/rspa.2012.0436}, methods exist for e.g., recursively computing the outbreak statistics\cite{ball_1986,ball_clancy_1993,miller2019distribution},
solving the master equation for the stochastic dynamics directly by numerical linear algebra\cite{doi:10.1098/rspa.2012.0436}, or deriving scaling laws for small outbreaks near threshold\cite{PhysRevE.69.050901,Scaling,PhysRevE.89.042108}. Yet, in addition to being numerically unstable for large populations, computationally expensive, or limited in scope, such methods also fail to provide physical and analytical insights into how unusual and extreme outbreaks occur.

Here we develop an analytical approach based on WKB-methods\cite{Assaf_2017,Dykman1994,Meerson2010} which provides a closed-form expression for the asymptotic outbreak distribution in SIR, SEIR, and COVID-19 models with fixed population sizes ($N$) and heterogeneity in infectivity and recovery\cite{Subramaniane2019716118,Covasim,Schwartz2020,Catching2020.08.12.20173047}. We show that each outbreak is described by a unique most-probable path, \textcolor{black}{ and provide an effective picture of how stochasticity is manifested during a given outbreak.} For instance, compared to the expected mean-field dynamics each outbreak entails a unique, depletion or boost in the pool of susceptibles and an increase or decrease in the effective recovery rate, depending on whether the final outbreak is larger or smaller than the mean-field prediction. Most importantly, unlike usual rare-event predictions for epidemic dynamics, such as extinction or other large fluctuations from an endemic state\cite{Assaf2010,Meerson2010,PhysRevLett.117.028302,Black_2011}, and fade-out\cite{PhysRevE.80.041130}, our results do not rely on metastability \cite{Nasell:Book,SchwartzJRS2011,Dykman1994,PhysRevE.77.061107,doi:10.1137/17M1142028}, and thus are valid for the comparatively short time scales of outbreaks, $\mathcal{O}(\ln{N})$ \cite{TURKYILMAZOGLU2021132902}. In sharp contrast to systems undergoing escape from a metastable state, we show that the outbreak distribution corresponds to an infinite number of distinct paths-- one for every possible extensive outbreak. Each outbreak connects two unique fixed-points in a Hamiltonian system, {\it both} with non-zero probability flux. \textcolor{black}{Hence, by solving a canonical problem in population dynamics and non-equilibrium statistical physics, we uncover a new degenerate class of rare events for discrete-state stochastic systems.}

\textit{Baseline model.} We begin with the Susceptible-Infected-Recovered (SIR) model, often used as a baseline model for disease outbreaks\cite{keeling:infectious_diseases,AnderssonBook,MathematicalepidemiologyPastPresenFuture,rodrigues2016application}. Individuals are either susceptible (capable of getting infected), infected, or recovered$/$deceased, \textcolor{black}{and can make transitions between these states through two basic processes: infection and recovery}. Denoting the total number of susceptibles $S$, infecteds $I$, and recovereds $R$ in a population of fixed size $N$, the probability per unit time that the number of susceptibles decreases by one and the number of infecteds increases by one is $\beta SI\!/\!N$, where $\beta$ is the infectious contact rate\cite{keeling:infectious_diseases,AnderssonBook,rodrigues2016application}. Similarly, the probability per unit time that the number of infecteds decreases by one is $\gamma I$, where $\gamma$ is the recovery rate\cite{keeling:infectious_diseases,AnderssonBook,rodrigues2016application}. \textcolor{black}{Combining both processes results in a discrete-state system with the following stochastic reactions:
\begin{eqnarray}
\label{eq:SIRreac1}&(S,I) \rightarrow (S-1,I+1) \;\; \text{with rate }\; \beta S I/N,\\
\label{eq:SIRreac2}&(I,R) \rightarrow (I-1,R+1) \;\; \text{with rate }\; \gamma I.
\end{eqnarray}}As $N$ is assumed constant, the model is appropriate for the short time scales of early emergent-disease outbreaks, for example, with an assumed separation between the outbreak dynamics and demographic time scales, as well as re-infection\cite{keeling:infectious_diseases}. From the basic reactions (\ref{eq:SIRreac1}-\ref{eq:SIRreac2}), the master equation describing the probability of having $S$ susceptibles and $I$ infecteds at time $t$ is
\begin{align}
\label{eq:M}
&\frac{\partial P}{\partial t} (S,I,t) = -\frac{\beta SI}{N}P(S,I,t) -\gamma IP(S,I,t) + \\
&\frac{\beta (S\!+\!1)(I\!-\!1)}{N}P(S\!+\!1,I\!-\!1,t) +\gamma(I\!+\!1)P(S,I\!+\!1,t)\nonumber.
\end{align}
\textcolor{black}{Solving this equation allows one to predict the probability that a particular proportion of a population eventually becomes infected for a given set of parameters. This is our goal here, as in many other works \cite{doi:10.1146/annurev-statistics-061120-034438,keeling:infectious_diseases,StochasticEpidemicModels,doi:10.1098/rspa.2012.0436,ball_1986,ball_clancy_1993,miller2019distribution, doi:10.1098/rspa.2012.0436,PhysRevE.69.050901,Scaling,PhysRevE.89.042108}.} Yet, in full generality such equations cannot be solved analytically, and one must resort to high-dimensional numerics, recursive computations, and$/$or large numbers of simulations\cite{doi:10.1098/rspa.2012.0436}. Yet, if $N$ is large it is possible to construct an asymptotic solution to Eq.(\ref{eq:M}) for all $\mathcal{O}(N)$ outbreaks using WKB methods\cite{Assaf_2017,Dykman1994,Meerson2010}, as we will show.

\textcolor{black}{First, to summarize what is known for large $N$, let us define the fraction of individuals in each disease state $x_{w}\!=\!W/N$ where $W\!\in\!\{S,I,R\}$. Note that as the total population size is constant, $1\!=\!x_{r}\!+\!x_{s}\!+\!x_{i}$. The mean-field limit of the reactions (\ref{eq:SIRreac1}-\ref{eq:SIRreac2}), corresponds to a simple set of differential equations: $\dot{x}_{s}=-\beta x_{i}x_{s}$, $\;\dot{x}_{i}=\beta x_{i}x_{s}-\gamma x_{i}$, and $\dot{x}_{r}=\gamma x_{i}$. Of particular interest is the total fraction of the population infected in the long-time limit, $x_{r}^{*}=x_{r}(t\!\rightarrow\!\infty)$, whose average, $\overline{x_{r}^{*}}$,  can be found by integrating the mean-field system. For small initial fractions infected, the solution (according to the mean-field) depends only on the basic reproductive number, $R_{0}\!\equiv\!\beta/\gamma$ \cite{keeling:infectious_diseases,AnderssonBook,MathematicalepidemiologyPastPresenFuture,rodrigues2016application}, and solves the equation $1-\overline{x_{r}^{*}}\!=\!e^{-R_0 \overline{x_{r}^{*}}}$~\cite{keeling:infectious_diseases,HARKO2014184}.}

\begin{figure}[h]
\center{\includegraphics[scale=0.24]{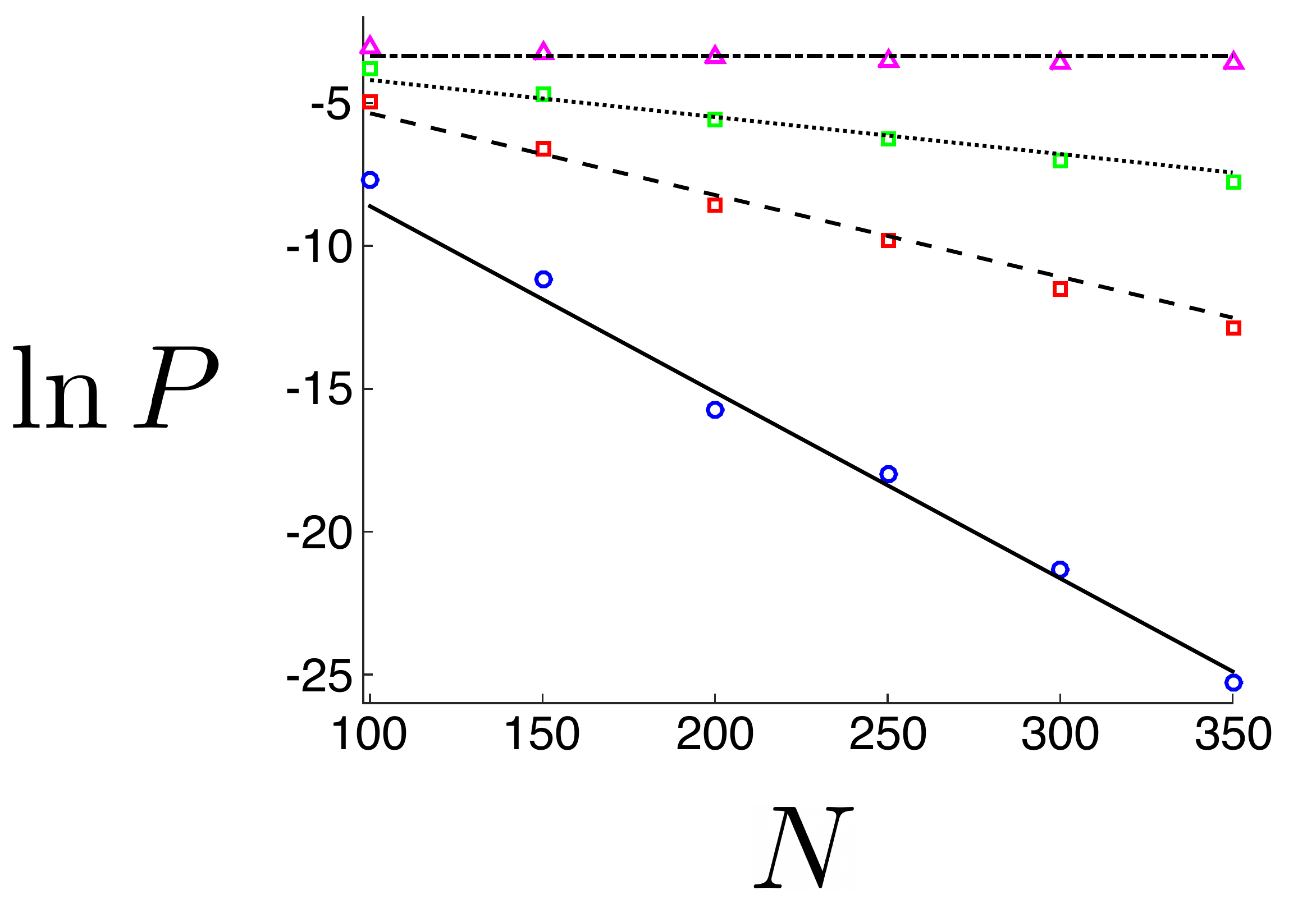}}
\vspace{-2mm}\caption{Extreme outbreak probability scaling with the population size in the SIR model. Plotted is the probability that $100\%$ (blue), $98\%$ (red), $96\%$ (green), \textcolor{black}{and $89\%$ (magenta)} of the population are infected during an outbreak vs $N$. Results from $10^{11}$ simulations (symbols) are compared with theoretical lines whose slopes are given by Eq.(\ref{eq:Action}). Here $R_{0}\!=\!2.5$.}
\label{fig:EO0}
\end{figure}

\textcolor{black}{But, what about a half, a fourth, twice, etc. of this expected outbreak, or a case in which the entire population eventually becomes infected? Since the SIR-model is inherently stochastic and governed by Eq.(\ref{eq:M}), such solutions are also possible. To get a sense of how the probabilities for various outbreaks arise, and to guide our analysis, we perform some stochastic simulations, and plot (on a semi-log scale) the fraction of outcomes that result in a given total-fraction infected. Examples are shown in Fig.\ref{fig:EO0} for outbreaks: $100\%$ (blue), $98\%$ (red), and $96\%$ (green) when $R_{0}\!=\!2.5$. For reference, the mean-field outbreak of $89\%$ \textcolor{black}{(magenta) is also plotted}. Here and throughout, simulations were performed using Gillespie's direct method\cite{Gillespie2013,keeling:infectious_diseases,doi:10.1146/annurev-statistics-061120-034438} starting from a single infectious individual. Notice that for each outbreak value, $\ln{P}$ is linear in $N$, with a slope that depends on the outbreak, $\ln{P}(x_{r}^{*})\!\simeq\!N\mathcal{S}(x_{r}^{*})$. This asymptotic WKB scaling is consistent with what we expect on general theoretical grounds for large deviations in stochastic population models with a small $\mathcal{O}(1/N)$ noise parameter \cite{Assaf_2017,Dykman1994,Assaf2010,Meerson2010}.}

\textcolor{black}{Equipped with the WKB hypothesis for the distribution of outbreaks, we substitute \textcolor{black}{the ansatz $P(x_{s},x_{i},t)\!\sim\!\exp[-N\mathcal{S}(x_{s},x_{i},t)]$ into Eq.(\ref{eq:M}), and keep leading-order terms in $N\gg 1$}. In particular, we do a Taylor expansion of $P(x_s,x_i,t)$; e.g., $\;P(x_s+ 1/N,x_i- 1/N,t)\simeq e^{-N \mathcal{S}(x_{s},x_{i},t)-\partial \mathcal{S}/\partial x_s+\partial \mathcal{S}/\partial x_i}$. This allows finding the leading-order solution~\footnote{\textcolor{black}{Sub-leading order contributions to the probability can be found by continuing the large-N expansion \cite{Assaf_2017,Assaf2010,Meerson2010,Black_2011}}}, called the action, given by $\mathcal{S}(x_s,x_i,t)$\cite{Assaf_2017,Assaf2010,Meerson2010}. Taking the large-$N$ limit in this way converts the master equation (\ref{eq:M}) into a Hamilton-Jacobi equation, $\partial_{t} {\cal S}(x_{s},x_{i},t)+H(x_{s},x_{i},p_{s},p_{i})\!=\!0$ \cite{Assaf_2017,Dykman1994}, with a Hamiltonian given by
\begin{equation}
H=\beta x_{i}x_{s}\big(e^{p_{i}-p_{s}}-1\big) +\gamma x_{i}\big(e^{-p_{i}}-1\big).
\label{eq:H}
\end{equation}
Here the momenta of the susceptibles and infecteds are respectively defined as $p_s=\partial\mathcal{S}/\partial x_s$ and $p_i=\partial\mathcal{S}/\partial x_i$.}

As a consequence, in the limit of $N\gg 1$ the outbreak dynamics satisfy Hamilton's equations: $\dot{x}_{w}\!=\!\partial H/\partial p_w$ and $\dot{p}_{w}\!=\!-\partial H/\partial x_w$, just as in analytical mechanics\cite{Landau1976Mechanics}. Furthermore, solutions are minimum-action\cite{Dykman1994}, or maximum-probability. Namely, given boundary conditions for an outbreak, Hamilton's equations will provide the most-likely dynamics. As in mechanics, once the dynamics are solved, the action ${\cal S}(x_{s},x_{i},t)$ can be calculated along an outbreak path:
\begin{equation}\label{action}
{\cal S}(x_{s},x_{i},\textcolor{black}{t})=\textcolor{black}{\int_{0}^{t}\!\!\left(p_{s}\dot{x}_{s}+p_{i}\dot{x}_{i}-H\right)dt'}
%\int p_{s} dx_{s}' +\int p_{i} dx_{i} -\int H dt.
\end{equation}

Before continuing our analysis, let us comment on the distribution, $P(x_{s},x_{i},t)$, and explain the sense in which certain outbreaks are extreme. As $P(x_{s},x_{i},t)$ scales exponentially with $N$ (for large $N$), if the action ${\cal S}(x_{s},x_{i},t)$ associated with an outbreak differs significantly from $0$, the outbreak will occur with an exponentially small probability, just as we observe in Fig.\ref{fig:EO0}. \textcolor{black}{In fact, the special case of ${\cal S}\!=\!0$ ($p_{i}\!=\!p_{s}\!=\!0$) is nothing other than the aforementioned mean-field prediction, which nicely quantifies why it is the most-likely extensive outbreak.}

\textcolor{black}{\textit{Results.} In order to find the probability distribution of outbreaks, we observe that Hamiltonian (\ref{eq:H}) does not depend explicitly on time; that is $H$ evaluated along an outbreak is conserved in time\cite{Landau1976Mechanics}. Now, we substitute $\dot{p}_{i}=-\partial H/\partial x_i$, and write the Hamiltonian for the SIR model in a suggestive form, $H\!=\!-x_{i}\;\dot{p}_{i}$. Thus, if we consider the same large-population limit as the usual mean-field analysis discussed above, and restrict ourselves to outbreaks that start from small infection, e.g., $x_{i}(t\!=\!0)\!=\!1/N$ with $N\!\gg\!1$, it must be that $H\!\simeq\!0$. \textcolor{black}{Notably, because the energy is zero, we can drop the explicit time dependence in Eq.(\ref{action})}. As a result, since the number of infecteds grows and then decreases during the course of an outbreak with $x_{i}(t)\!\neq\!0$ for general $t$, one must have $p_{i}\!=\!$ {\it const.}}

%In order to find the probability distribution of outbreak sizes and outbreak paths, we note the following observations: First, the Hamiltonian (\ref{eq:H}) does not depend explicitly on time; that is $H$ evaluated along an outbreak is conserved in time, just as in mechanics\cite{Landau1976Mechanics}. Second, in order to more precisely understand the conserved energy, we can write the Hamiltonian for the SIR model in the suggestive form, $H\!=\!-x_{i}\;\dot{p}_{i}$, which follows from direct substitution of $\dot{p}_{i}=-\partial H/\partial x_i$. Third, if we consider the same large-population limit as the usual mean-field analysis discussed above, and restrict ourselves to outbreaks that start from small infection, e.g., $x_{i}(t\!=\!0)\!=\!1/N$ with $N\!\gg\!1$, it must be that $H\!\simeq\!0$. \textcolor{black}{Because the energy is a constant (zero), we can drop the explicit time dependence in Eq.(\ref{action})}. Fourth, since the number of infecteds grows and then decreases during the course of an outbreak with $x_{i}(t)\!\neq\!0$ for general $t$, one must have $p_{i}\!=\!$ {\it const.}}

\textcolor{black}{At this point, we highlight a crucial difference between our analysis for stochastic outbreak dynamics, and the traditional use of WKB for analyzing large deviations in population models with metastable states. In the latter, the traditional $H\!\simeq\!0$ condition of the WKB usually derives from the fact that the model has a locally unique stable fixed-point for the mean-field coordinates, e.g, $\dot{\bold{x}}\!=\!0$ \cite{Assaf2010,Meerson2010,PhysRevLett.117.028302,Black_2011, Nasell:Book,SchwartzJRS2011,Dykman1994,PhysRevE.77.061107,doi:10.1137/17M1142028}. Common examples are stochastic switching and extinction from endemic equilibria. In our case, the zero-energy condition corresponds to a conserved momentum, and in fact, an infinite number of them.
The non-zero momentum boundary conditions entailed by the conserved momenta are distinct from other known categories of extreme processes in discrete-state non-equilibrium systems and stochastic populations, and hence we uncover a new {\it degenerate} class.}

\textcolor{black}{Now that we know that outbreaks in the SIR model are defined according to a \textcolor{black}{conserved} momenta, i.e., $m\equiv e^{p_i}$, we can equate the Hamiltonian~(\ref{eq:H}) to zero,
and find the non-constant fluctuational momentum $p_{s}$, along an outbreak in terms of $x_{s}$, $m$, and $R_{0}$,
\begin{equation}
p_{s}= \ln\left\{R_{0}x_{s}m^{2}/[m(R_{0}x_{s}+1)-1]\right\}.
\label{eq:ps}
\end{equation}
This momentum is necessary for evaluating Eq.~(\ref{action}). Continuing on toward our main goal of calculating the action, we note that the integral over $p_i$ vanishes, since it is a constant of motion and $x_i(t=0)=x_i(t\to\infty)\simeq 0$. Furthermore, the integral over $H$ also vanishes since $H\simeq 0$. As a result, in order to determine the action, we need to compute the integral over $p_s$ [Eq.~(\ref{eq:ps})] from the initial state $x_s=1$ to the final state $x_{s}(t\!\rightarrow\!\infty)=x_s^*$. The only thing left for us is to express  $x_s^*$ in terms of the \textcolor{black}{conserved momentum} $m$. \textcolor{black}{This can be done by using Hamilton's equations, see SM for details}. Doing so, we arrive at the total action accumulated in the course of an outbreak
\begin{align}
\label{eq:Action}
&{\cal S}(\textcolor{black}{x_{s}^{*}})=\ln x_s^*+(1-x_s^*)\\
&\times\left[m(1+R_0x_s^*)-1+\ln\left[(m(R_0+1)-1)/(x_s^* m^2 R_0)\right]\right]\nonumber.
\end{align}
\noindent \textcolor{black}{Note that $\mathcal{S}$ is a function of $x_s^*$ only, since for fixed $R_{0}$ there is a complete mapping between the final outbreak size and $m$ (see SM Eq.(A9) for $x_s^*(m)$).} Equation (\ref{eq:Action}) is our main result: the asymptotic solution of Eq.(\ref{eq:M}) for the distribution of all ${\cal O}(N)$ outbreaks.~\footnote{\textcolor{black}{For brevity we have dropped the dependence on $x_{i}$ in the final-outbreak action Eq.(\ref{eq:Action}), since all final states have the same $x_{i}\!=\!0$}}}
%when $N$ is large but finite.}

\begin{figure}[t]
\center{\includegraphics[scale=0.23]{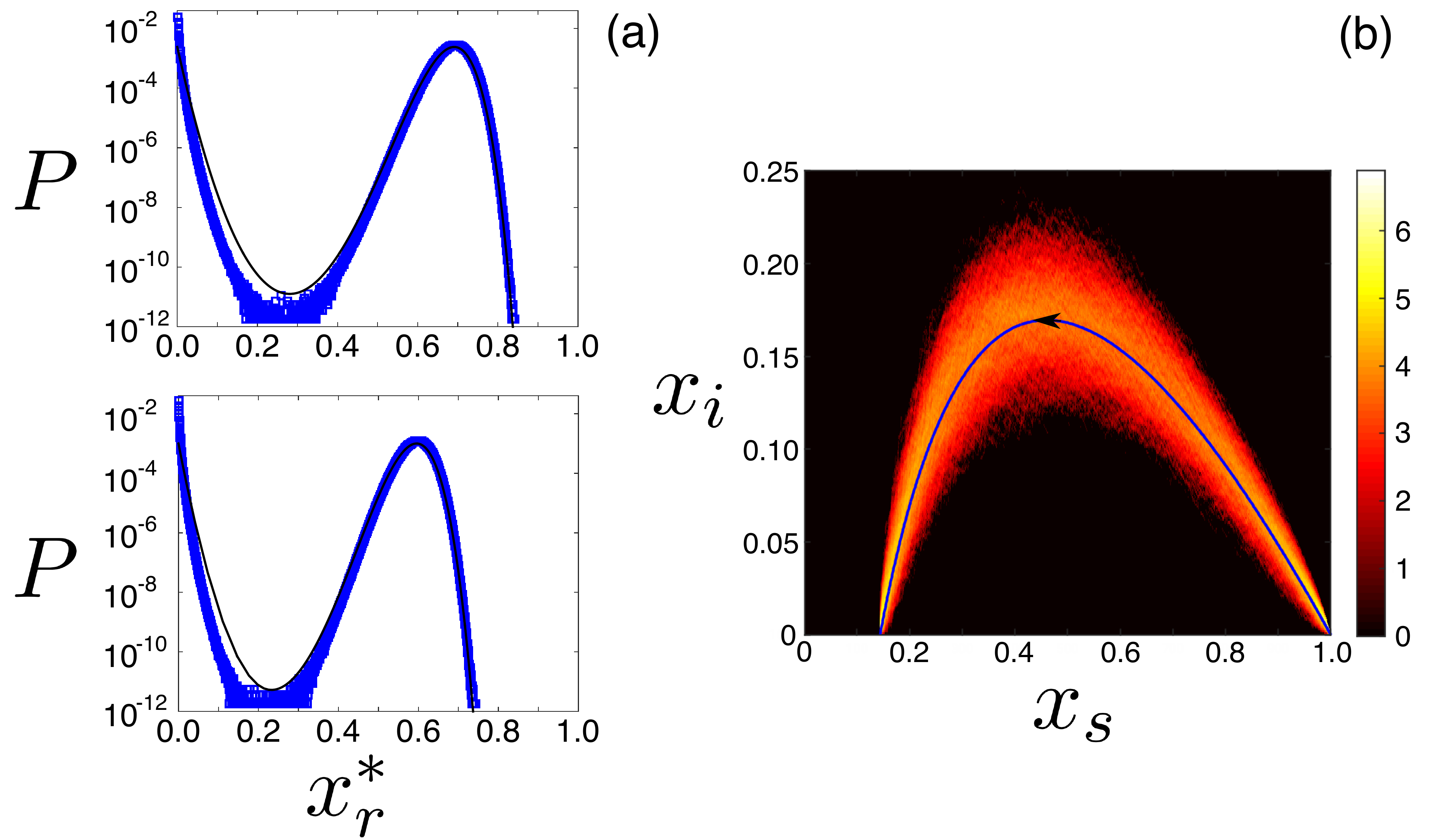}}
\vspace{-5mm}\caption{Outbreak distributions. (a) Final outbreak distribution for the SIR model (top) and a higher-dimensional COVID-19 model (bottom). Stochastic simulation results (blue squares) are compared with theory (black lines). Parameters are given in the main text. \textcolor{black}{Despite the varying complexity, the outbreak distributions in both models are captured by the same theory.}
%The expected outbreak size (mean-field solution) is obtained approximately at $x_r^*\simeq 0.69$.
(b) \textcolor{black}{Histogram of 2000 stochastic trajectories in the SIR model that result in the same final \textcolor{black}{(non mean-field)} outbreak} $x_{r}^{*}\!=\!0.86$. The Eq.(\ref{eq:xi}) prediction is shown with a blue curve. Parameters are $N\!=\!1000$ and $R_{0}\!=\!1.7$. The colormap for the histogram is on log-scale.}
\label{fig:EO1}
\end{figure}

\textcolor{black}{Our main result can now be tested in several ways. First, we go back to the motivating Fig.\ref{fig:EO0}. Recall that our approach predicts that, as a function of $N$, the action gives the slope of $\ln{P}(x_{r}^{*})\!\simeq\!N\mathcal{S}(x_{r}^{*})$. As such, we can overlay lines in Fig.\ref{fig:EO0}, where the slopes are predictions from Eq.(\ref{eq:Action}). Doing so for three extreme outbreak values (as well as the mean-field), we observe very good agreement, especially for larger values of $N$. Second, we can fix $N$ and $R_{0}$, and see how well Eq.(\ref{eq:Action}) predicts the full distribution. Such comparisons with stochastic simulations are shown in the upper panel of Fig.\ref{fig:EO1} (a).} In particular, we plot the fraction of $10^{12}$ simulations that resulted in an outbreak $x_{r}^{*}$ in blue, and the solutions of Eq.~(\ref{eq:Action}) with a black line. Again, the agreement between the two is quite good for the population size $N\!=\!2000$ and $R_{0}\!=\!1.7$. Disagreement increases as the outbreak sizes approach $\mathcal{O}(1/N)$. \textcolor{black}{Qualitatively, we can see that our theory captures the full cubic structure of the outbreak distribution, with a local maxima at the smallest outbreak (here $1/N$) and the mean-field solution, $\overline{x_r^*}\simeq 0.69$~\cite{Assaf_2017,Assaf2010,SchwartzJRS2011,PhysRevE.80.041130}}.

{\color{black} To get more insight into the outbreak distribution, one can use Eq.~(\ref{eq:Action}) to compute the action, e.g. in the vicinity
of the mean-field, $\overline{x_r^{*}}$. Locally the distribution is a Gaussian around $\overline{x_r^{*}}$, with a relative variance that takes a minimum at $R_0\simeq 5/3$, for which stochastic deviations from the mean-field outbreak are minimized (See SM for further details on the distribution's unique shape).}
%{\color{cyan}To get more insight into the outbreak distribution, we can use Eq.~(\ref{eq:Action}) to compute the action in the vicinity of the mean-field outbreak $\overline{x_r^{*}}$. Denoting $\overline{x_s^*} =1-\overline{x_r^{*}}$, we find: ${\cal S}(x_s^*)\simeq (1/2){\cal S}''(\overline{x_s^*})(x_s^*-\overline{x_s^*})^2$, with ${\cal S}''(\overline{x_s^*})=(R_0\overline{x_s^*}-1)^2/[(1-\overline{x_s^*})\overline{x_s^*}(R_0^2\overline{x_s^*}+1)]$. This means that close to its maximum the outbreak distribution is a Gaussian with a width of $\sigma=(N {\cal S}''(\overline{x_s^*}))^{-1/2}$. Moreover, we find  that the distribution's coefficient of variation (COV), $COV=\sigma/\overline{x_s^*}$, receives a minimum at $R_0=1.66$. That is, the deviation from the mean-field outbreak size is minimized at $R_0\simeq 5/3$, see Fig. S? in the SM. }
%In between is a minimum (near $x_{r}^{*}\!\approx\!0.28$ in Fig.\ref{fig:EO1} (a)) that we call the least-likely small outbreak, $x_{r}^{\text{m}}$. This quantity can be computed by solving $\partial {\cal S}/\partial x_{s}^{*}(x_{r}^{\text{m}})=0$ using Eq.~(\ref{eq:Action}) For outbreaks smaller than the mean-field solution, we can use $x_{r}^{\text{m}}$ to separate outbreaks into increasing or decreasing likelihoods.
%The least-likely small outbreak gives us another novel observable of our approach, and one that can be parametrically tested as a function of $R_{0}$. An example is shown in Fig.\ref{fig:EO1} (b), where we plot...

Before moving to more general outbreak models we mention a few important qualitative details that emerge from our approach. In particular, let us consider the stochastic dynamics for the fraction of the population infected, $\dot{x}_i=\partial H/\partial p_i$. Substituting Eq.(\ref{eq:ps}) into $\dot{x}_i$, yields
\begin{equation}
\label{eq:Infectious}
\dot{x}_{i}=\beta x_{i}\Big[(m-1)/(m R_{0}) +x_{s}\Big] -(\gamma/m)x_{i}.
\end{equation}
First, note that when $m\!=\!1$ ($p_{i}\!=\!p_{s}\!=\!0$), we uncover the mean-field SIR model system, $\dot{x}_{i}=\beta x_{i}x_{s} -\gamma x_{i}$. From the mean-field,
we can recover Eq.(\ref{eq:Infectious}) with the suggestive transformations $x_{s}\!\rightarrow\! x_{s} +(m-1)/mR_{0}$, and $\gamma\rightarrow\gamma/m$~\footnote{A similar effect occurs in cell biology in a mRNA-protein genetic circuit, where fluctuations in the mRNA copy number can be effectively accounted for by taking a protein-only model
with a modified production rate~\cite{roberts2015dynamics}.}. Recalling that each outbreak is parameterized by a unique constant $m$, evidently the effect of demographic stochasticity is to add an effective constant reduction (or boost) to the pool of susceptibles and to increase (or decrease) the effective recovery rate, depending on whether the final outbreak is smaller ($m<1$) or larger ($m>1$) than the mean-field, respectively.

\textcolor{black}{We can test our prediction that a conserved $m$ constrains an entire outbreak path by picking a particular final outbreak size, corresponding to a particular value of $m$, and compare to stochastic trajectories. One method for comparison is to build a histogram in the $(x_{i},x_{s})$ plane from many simulations that end in the same outbreak size, and plot the constant-$m$ prediction. The latter can be found by solving the differential equation $dx_{i}/dx_{s}=\dot{x}_{i}/\dot{x}_{s}$} from Hamilton's equations, or
\begin{equation}
x_{i}(x_{s},m)= 1-x_{s} +\ln\!\left[\!\frac{m(R_{0}x_{s}+1)-1}{m(R_{0}+1)-1}\!\right]\big/\!R_{0}m.
\label{eq:xi}
\end{equation} An example is shown in Fig.\ref{fig:EO1} (b) for a final outbreak of $86\%$ when $R_{0}=1.7$ (the mean-field prediction is $69\%$). \textcolor{black}{The color map for the histogram is plotted along with the prediction from Eq.(\ref{eq:xi}). As expected, the outbreak-path prediction lies in the maximum density region. Thus, not only does our approach predict probabilities, but also the optimal dynamics that leads to outbreaks-- driven by an effective conserved momentum, $m$.}

\textcolor{black}{\textit{General model.} We now generalize our results to more complex and realistic outbreak models. Typically, such models derive from the same basic assumptions as SIR, but have more states and free parameters.} \textcolor{black}{For example}, epidemiological predictions for COVID-19 (at a minimum) require an incubation period of around $5$ days, and an asymptomatic disease state, i.e., a group of people capable of spreading the disease without documented symptoms\cite{Subramaniane2019716118,Covasim,Schwartz2020}. \textcolor{black}{Both features: finite incubation and heterogeneity in infectious states, can form the basis of a more general class of outbreak models\cite{ModelingCOVID-19,Ray2020.08.19.20177493,Subramaniane2019716118,Covasim,Schwartz2020,Catching2020.08.12.20173047}}. Within this class, we assume that upon infection, susceptible individuals first become exposed (E), and then enter an infectious state at a finite rate $\alpha$. By assumption there are several possible infectious states (e.g., asymptomatic, mild, severe, tested, quarantined, etc.) that an exposed individual can enter according to prescribed probabilities~\cite{ModelingCOVID-19,Ray2020.08.19.20177493,Subramaniane2019716118,Covasim,Schwartz2020,Catching2020.08.12.20173047}. In addition, infectious states can have their own characteristic infection rates and recovery times. Putting these ingredients together, let us define $\mathcal{N}$ infectious states, $I_{n}$, where $n\!\in\!\{1,2,...,\mathcal{N}\}$, each with their own infectious contact rate $\beta_{n}$ and recovery rate $\gamma_{n}$, and which appear from the exposed state with probabilities $z_{n}$\cite{Schwartz2020,Catching2020.08.12.20173047,10.1371/journal.pone.0244706,Subramaniane2019716118,Covasim}. \textcolor{black}{See SM for list of reactions.}

Following the WKB-prescription above, the Hamiltonian \textcolor{black}{for our general class of outbreak models} is
\begin{eqnarray}\label{eq:H2}
H=\sum_{n}&\beta_{n} x_{i,n}x_{s}\big(e^{p_{e}-p_{s}}-\!1\big) +\gamma_{n} x_{i,n}\big(e^{-p_{i,n}}-\!1\big) \nonumber \\
+&\!\!\!\!\!\!\!\!\!\!\!\!\!\!\!\!\!\!\!\!\!\!\!\!\!\!\!\!\!\!\!\!\!\!\!\!\!\!\!\!\!\!\!\!\!\!\!\!\!\!\alpha z_{n}x_{e}\big(e^{p_{i,n}-p_{e}}-\!1\big).
\end{eqnarray}
\textcolor{black}{Despite the increased dimensionality and parameter heterogeneity, the general outbreak system defined through Eq.(\ref{eq:H2}) can also be solved analytically by precisely the same approach as the baseline SIR model. As in the latter, the essential property that makes the system solvable is the constancy of all momenta except for $p_{s}$. This property ensures that, here again, there is one free constant, $m$, that determines all momenta and the final outbreak size. Demonstrating this requires a few additional steps of algebra, but the result is a simple update to Eq.(\ref{eq:Action}) that involves a sum over the heterogeneities $\{z_{n},\beta_{n},\gamma_{n}\}$. See SM for general outbreak solution, Eq.(A29). An important consequence of the general solution is that, in the special case of the SEIR model\cite{keeling:infectious_diseases}, where there is only one infectious state, the outbreak action is identical to the SIR model, Eq.(\ref{eq:Action}). Namely, finite incubation changes the dynamics of outbreaks, but has only a sub-exponential contribution to their probability.}

An example prediction from our general analysis is shown in the lower panel of Fig.\ref{fig:EO1} (a). The analytical solution (black line) is in very good agreement \textcolor{black}{with stochastic simulations of a COVID-19 model with asymptomatic ($n\!=\!1$) and symptomatic ($n\!=\!2$) infectious individuals. The infection parameters\footnote{For COVID-19 modelling, a typical choice for time units would be $t\!=\!1$ corresponding to 10 days.} take realistic heterogeneous values, i.e., $\beta_{1}\!=\!1.8$, $\beta_{2}\!=\!1.12$, $\gamma_{1}\!=\!1$, $\gamma_{2}\!=\!0.8$, $\alpha\!=\!2$, $z_{1}\!=\!0.3$, and $N\!=\!4000$ \cite{Schwartz2020,Catching2020.08.12.20173047,10.1371/journal.pone.0244706,Subramaniane2019716118,Covasim}, where $z_{1}=0.3$ is a typical value for the fraction of asymptomatic infection.} \textcolor{black}{Despite the increased complexity, the distribution in the more general model is also well-captured by our theory.}

Before concluding, it is worth mentioning that although in real outbreaks the parameters in Eq.(\ref{eq:H2}) may fluctuate in time, if the fluctuations are fast compared to outbreak time-scales $\mathcal{O}(\ln{N})$ \cite{TURKYILMAZOGLU2021132902}, we expect the distribution to approach the SIR model with effective time-averaged parameters, which can be computed using methods detailed in~\cite{assaf2008population, assaf2013extrinsic}. On the other hand, if the fluctuations are slow with respect to the same time scales, we expect the distribution to be described by integrating over the solution of Eq.(\ref{eq:H2}), with weights given by the probability-density of rates~\cite{assaf2008population, assaf2013extrinsic}. In the intermediate regime, one must solve a Hamiltonian system with increased dimensionality, which includes both demographic noise and environmental variability. In this way, our results can provide a basis for understanding even more realistic outbreak dynamics.

\textit{Conclusions.} \textcolor{black}{We solved the canonical problem of predicting the outbreak distribution of epidemics in large, fixed-sized populations.}
Our theory was based on the exponential scaling of the probability of extensive outbreaks on the population size, which allowed the use of a semiclassical approximation. \textcolor{black}{By analyzing SIR, SEIR, and COVID-19 models, we were able to derive simple formulas for the paths and probabilities of all extensive outbreaks, and find an effective picture of how stochasticity is manifested during outbreaks.} Most importantly we showed that, unlike other well-known examples of rare events in population models, the statistics of extreme outbreaks depend on an infinite number of minimum-action paths satisfying \textcolor{black}{a unique set of boundary conditions with conserved momenta}. Due to their distinct and degenerate phase-space topology, extreme outbreaks represent a new class of rare process for discrete-state stochastic systems. \textcolor{black}{As with other extreme processes, our solution can form the basis for predictions in many other scenarios, including stochastic outbreaks mediated through complex networks.}

JH and IBS were supported by the U.S. Naval Research Laboratory funding
(N0001419WX00055), and the Office of Naval Research (N0001419WX01166) and
(N0001419WX01322). MA was supported by the Israel Science Foundation Grant No. 531/20, and by the Humboldt Research Fellowship for Experienced Researchers of the Alexander von Humboldt Foundation.

\section*{Appendix (Supplementary Material)}
\renewcommand{\theequation}{A\arabic{equation}}
\renewcommand{\thefigure}{S\arabic{figure}}
\setcounter{figure}{0}
\setcounter{equation}{0}

\section{\label{SM:Sec1} SIR model: WKB approximation and outbreak distribution}
In this section we use the WKB approximation to obtain Hamilton's equations governing the (leading order of the) stochastic dynamics in the SIR model.
Then we derive the probability distribution of outbreak sizes and the trajectory of infected along an outbreak.

To treat the master equation [Eq.~(3)] via the WKB method we employ the ansatz $P(x_{s},x_{i})\sim e^{-N\mathcal{S}(x_{s},x_{i})}$, where $\mathcal{S}$ is called the action function, and $x_s=S/N$ and $x_i=I/N$ are the fractions of susceptibles and infected respectively. Substituting the WKB ansatz into Eq. (3), we  do a Taylor expansion of the action around $(x_s,x_i)$; i.e., $\;P(x_s+ 1/N,x_i- 1/N)\simeq e^{-N \mathcal{S}(x_{s},x_{i})-\partial \mathcal{S}/\partial x_s+\partial \mathcal{S}/\partial x_i}$
and $\;P(x_s,x_i+ 1/N)\simeq e^{-N \mathcal{S}(x_{s},x_{i})+\partial \mathcal{S}/\partial x_i}$. Doing so, we arrive at a Hamilton-Jacobi equation, $\partial_{t} {\cal S}(x_{s},x_{i})+H(x_{s},x_{i},p_{s},p_{i})\!=\!0$, where the Hamiltonian is given by [Eq.(4)], or
\begin{equation}
H=\beta x_{i}x_{s}\left(e^{p_{i}-p_{s}}-1\right) +\gamma x_{i}\left(e^{-p_{i}}-1\right).
\label{SM:Hamil}
\end{equation}
Here the momenta of the susceptibles and infected are respectively defined as $p_s=\partial\mathcal{S}/\partial x_s$ and $p_i=\partial\mathcal{S}/\partial x_i$.

As in classical mechanics, the outbreak dynamics satisfy Hamilton's equations
\begin{align}
\label{H1}&\dot{x}_s=\frac{\partial H}{\partial p_s}=-\beta x_i x_s e^{p_i-p_s},\\
\label{H2}&\dot{x}_i=\frac{\partial H}{\partial p_i}=\beta x_i x_s e^{p_i-p_s}-\gamma x_i e^{-p_i},\\
\label{H3}&\dot{p}_s=-\frac{\partial H}{\partial x_s}=-\beta x_i (e^{p_i-p_s}-1),\\
\label{H4}&\dot{p}_i=-\frac{\partial H}{\partial x_i}=-\beta x_s (e^{p_i-p_s}-1)-\gamma(e^{-p_i}-1).
\end{align}

As shown in the main text,  since the Hamiltonian does not depend explicitly on time it is a constant of motion, and furthermore, it can be shown that $H\simeq 0$ when $x_{i}(t=0)\simeq 0$. As a consequence, $p_i$ is a free constant throughout the epidemic outbreak, and we can define the constant $m=e^{p_i}$. Equating the Hamiltonian (\ref{SM:Hamil}) to zero, after some algebra we find:
\begin{align}
&e^{p_{s}}= R_{0}x_{s}m^{2}/[m(R_{0}x_{s}+1)-1].
\label{SM:ps}
\end{align}
At this point we can use Hamilton's equations (\ref{H1}-\ref{H2}) to express the final outbreak size in
terms of the initial momentum of infected. Because the total population density is constant, the density of the recovered individuals, $x_r=R/N$, satisfies: $\dot{x}_r=-\dot{x}_s-\dot{x}_i=\gamma x_i/m$.
As a result, we can write a differential equation for $dx_s/dx_r$ by dividing Eq.~(\ref{H1}) by $\dot{x}_r$, which yields:
\begin{equation}
\frac{dx_s}{dx_r}=-R_0 x_s m^2 e^{p_s}.
\end{equation}
Substituting $e^{p_s}$ from Eq.~(\ref{SM:ps}) into this equation, and integrating $x_{s}$ from $1$ to $x_{s}^{*}$, and $x_{r}$ from $0$ to $1-x_{s}^{*}$,  we find an implicit equation for the final outbreak versus $m$:
\begin{equation}
e^{R_{0}m(1-x_{s}^{*})}=[m(R_{0}+1)-1]/[m(R_{0}x_{s}^{*}+1)-1].
\label{SM:xs_star}
\end{equation}
This equation can be solved for $m$, and the result is
\begin{equation}\label{SM:formalsol}
x_s^*=\left\{1\!-\!m\!-\!W_0\left[(1\!-\!m(R_0\!-\!1))e^{1-m(R_0\!+\!1)}\right]\right\}/(mR_0),
\end{equation}
where $W_0(z)$ is the principle solution for $w$ of $z=we^w$.
Thus, for fixed $R_{0}$, we have a complete mapping between the final outbreaks and the free-parameter $m$: each outbreak corresponds to a unique value of $m$.

In addition to finding the final outbreak size as function of $m$, we can find the trajectory for the population fraction infected during an outbreak by solving $dx_{i}/dx_{s}$. Using Hamilton's equations (\ref{H1}-\ref{H2}) and Eq.~(\ref{SM:ps}) we find
\begin{equation}
\frac{dx_i}{dx_s}=-1+\frac{1}{m(R_0x_s+1)-1}.
\end{equation}
Integrating $x_{i}$ from $0$ to $x_i$, and $x_{s}$ from $1$ to $x_s$,  results in Eq.(9) from the main text, or
\begin{align}
&x_{i}(x_{s},m)= 1-x_{s} +\ln\!\left[\!\frac{m(R_{0}x_{s}+1)-1}{m(R_{0}+1)-1}\!\right]\big/\!R_{0}m.
\label{SM:xi_xs}
\end{align}
Notably, at $m=1$ (the mean-field solution), Eq.~(\ref{SM:xs_star}) becomes
$x_s^*=e^{-R_0(1-x_s^*)}$, which yields the well-known mean-field total outbreak size
$x_r^*=1-x_s^*=1+W_0\left(-R_0e^{-R_0}\right)/R_0$. In addition, Eq.~(\ref{SM:xi_xs}) becomes
$x_i(x_s)=1-x_s+\ln(x_s)/R_0$, and we recover the well-known mean-field result of how the fraction of infected depends on that of the susceptibles.

Having found $p_{s}(x_{s})$, and $x_s^*(m)$ we can find the action by integrating $\int p_{s} dx_{s}$ between $x_s(0)=1$ and $x_s^*$. The result is Eq.~(7) from the main text.

\section{Generalized SIR model}
In this section, we generalize the SIR-model results to a broader class of outbreak models with finite incubation and heterogeneity in infectious states. First, let us list the possible reactions in the larger class (typical of COVID-19 models) described in the main text:
\begin{align}
\label{eq:Reactions}
&(S,E) \rightarrow (S\!-\!1,E\!+\!1) \;\; \text{with rate }\; S\sum_{n}\beta_{n} I_{n}/N, \\
&(E,I_{n}) \rightarrow (E\!-\!1,I_{n}\!+\!1) \;\; \text{with rate }\; z_{n}\alpha E, \\
&(I_{n},R) \rightarrow (I_{n}\!-\!1,R\!+\!1) \;\; \text{with rate }\; \gamma_{n} I_{n}.
\end{align}
From these, the Hamiltonian Eq.~(10) directly follows from the WKB limit described in the main text and in SM.\ref{SM:Sec1}:
\begin{eqnarray}\label{eq:H2A}
H&=&\sum_{n}\beta_{n} x_{i,n}x_{s}\left(e^{p_{e}-p_{s}}-\!1\right)  \\
&+&\alpha z_{n}x_{e}\left(e^{p_{i,n}-p_{e}}-\!1\right)+\gamma_{n} x_{i,n}\left(e^{-p_{i,n}}-\!1\right).\nonumber
\end{eqnarray}

To solve the system Eq.(\ref{eq:H2A}), let us adopt the convenient notation, $m\!=\!e^{p_{e}}$, $m_{s}\!=\!e^{p_{s}}$, and $m_{i,n}\!=\!e^{p_{i,n}}$. As before, we look for solutions with $\dot{p}_{i,n}=-\partial H/\partial x_{i,n}=0$, which implies
\begin{equation}
\gamma_{n}(1-1/m_{i,n})/\beta_{n}\!=\!x_{s}(m/m_{s}-1).
\label{eq:mi}
\end{equation}
Because the left hand side is a constant and the right hand side has no explicit dependence on ($\beta_{n}$,$\gamma_{n}$), we define an outbreak {\it constant},
\begin{equation}
\label{eq:C1}
C(m)\equiv x_{s}(m/m_{s}-1).
\end{equation}

Second, because $H\!=\!-x_{e}\dot{p}_{e}\!-\!\sum_{n}x_{i,n}\dot{p}_{i,n}=0$, by substitution of Hamilton's equations into Eq.(\ref{eq:H2A}), we have that $\dot{p}_{e}\!=\!0$. From $\dot{p}_{e}=-\partial H/\partial x_{e}=0$, we get
\begin{equation}
m=\sum_{n}z_{n}m_{i,n}.
\label{eq:me}
\end{equation}
Combining Eqs.(\ref{eq:mi}-\ref{eq:me}), we find that $C(m)$ is a constant solution of
\begin{equation}
\label{eq:C}
m=\sum_{n}\frac{z_{n}}{1-C(m)\beta_{n}/\gamma_{n}}.
\end{equation}
So far, we have a free constant $m$, which determines $C$ and $m_{i,n}$.

Next, in order to calculate the general outbreak action,
\begin{equation}\label{action2}
{\cal S}(\bold{x})=\int\!p_{s} dx_{s} +\int\!p_{e} dx_{e} + \sum_{n}\int\!p_{i,n} dx_{i,n} -\int\!H dt,
\end{equation}
we need to know the upper and lower limits for the integrals in Eq.(\ref{action2}). As with the SIR model, since: all momenta are constant except $p_{s}$, $H=0$, $x_{e}(t\!=\!0)\!\approx\!0$, $x_{i,n}(t\!=\!0)\!\approx\!0$, $x_{e}(t\!\rightarrow\!\infty)\!\rightarrow\!0$,
and $x_{i,n}(t\!\rightarrow\!\infty)\!\rightarrow\!0$, the only non-zero integral comes from $p_{s}$, which depends on $x_{s}(t\!\rightarrow\!\infty)\equiv x_{s}^{*}$.

One useful strategy for finding the final fraction of susceptibles $x_{s}^{*}$ is to find relationships between the time-integrals of $x_{e}$, $x_{i,n}$ , and $x_{s}$. As in the SIR model, let us start with three of Hamilton's equations determined from Eq.(\ref{eq:H2A}):
\begin{align}
\label{eq:Hamilton1}
&\dot{x}_{s} = -x_{s}\left(\frac{m}{m_{s}}\right)\sum_{n}\beta_{n}x_{i,n},\\
\label{eq:Hamilton2}
&\dot{x}_{i,n} = \alpha x_{e}z_{n}\left(\frac{m_{i,n}}{m}\right)-(\gamma_{n}/m_{i,n})x_{i,n},\\
\label{eq:Hamilton3}
&\dot{x}_{r} = \sum_{n}(\gamma_{n}/m_{i,n})x_{i,n}.
\end{align}
By defining $I_{e}\equiv\int_{0}^{\infty}\!x_{e}(t)dt$ and $I_{i,n}\equiv\int_{0}^{\infty}\!x_{i,n}(t)dt$, we can integrate Eqs.(\ref{eq:Hamilton2}-\ref{eq:Hamilton3}) with respect to $t$. Remembering that
$x_{i,n}(t=0)\!\approx\!0$,  $x_{i,n}(t\!\rightarrow\!\infty)\!\rightarrow\!0$, and  $x_{r}(t\!\rightarrow\!\infty)\!\rightarrow\!1-x_{s,}^{*}$, the result is:
\begin{align}
\label{eq:I2}
&0 = \alpha I_{e}z_{n}\left(\frac{m_{i,n}}{m}\right)-(\gamma_{n}/m_{i,n})I_{i,n}\\
\label{eq:I3}
&1-x_{s}^{*} = \sum_{n}(\gamma_{n}/m_{i,n})I_{i,n}.
\end{align}
Similarly, separating $t$ and $x_{s}$ in Eq.~(\ref{eq:Hamilton1}) and integrating over all time we get
\begin{align}
\label{eq:I1}
&\int_{1}^{x_{s}^{*}}\frac{m_{s}(x_{s})dx_{s}}{m\;x_{s}} =-\sum_{n}\beta_{n}I_{i,n}.
\end{align}
Finally, if we insert $m_{s}\!=\!mx_{s}/[x_{s}+C(m)]$ from Eq.(\ref{eq:C1}) into Eq.~(\ref{eq:I1}), we can solve Eqs.(\ref{eq:I2}-\ref{eq:I1}) for $x_{s}^{*}(m,C(m))$:
\begin{align}
\label{eq:FinalNew}
\frac{-\ln\!\left[\frac{x_{s}^{*}+C(m)}{1+C(m)}\right]}{1-x_{s}^{*}}=
\sum_{n} \frac{z_{n}}{m}\frac{\beta_{n}}{\gamma_{n}}\Bigg(\frac{\gamma_{n}}{\gamma_{n}-C(m)\beta_{n}}\Bigg)^{2}.
\end{align}

Using the constants $C(m)$ and $x_{s}^{*}$, the total accumulated action of an outbreak from Eq.(\ref{action2}) is given by the solvable integral
\begin{align}
\label{eq:LastIntegral}
{\cal S}(m,x_{s}^{*}(m,C(m))\!=\!\int_{1}^{x_{s}^{*}}\ln[m x_{s}/(x_{s}+C(m))]dx_{s},
\end{align}
and can be expressed as a function of the constant momentum $m$,
\begin{align}
\label{eq:FinalAction}
{\cal S}(m,x_{s}^{*}(m,C(m))= x_{s}^{*}\ln\left[\frac{mx_{s}^{*}}{C(m)+x_{s}^{*}}\right]\nonumber \\
-\ln\left[\frac{m}{C(m)+1}\right] +C(m)\ln\left[\frac{C(m)+1}{C(m)+x_{s}^{*}}\right].
\end{align}
All parameters in Eq.~(\ref{eq:FinalAction}) depend on $m$: $C(m)$ through Eq.~(\ref{eq:C}) and $x_{s}^{*}(m,C(m))$ through Eq.~(\ref{eq:FinalNew}).

\section{Shape of the outbreak distribution}
In this section we explore the unique shape of the outbreak distribution. In terms of analytical scaling, in Eq.~(\ref{SM:formalsol}) we have expressed $x_s^*$ (in the SIR model) as a function of $m$ which allows finding an explicit solution for the outbreak  distribution as a function of $x_s^*$. While this gives rise to a cumbersome expression, further analytical progress can be done in the vicinity of $m\simeq 1$, which is a local maximum of the distribution. Expanding the right hand side of Eq.~(\ref{SM:formalsol}) around $m=1$, we find $m$ in terms of $x_s^*$,
\begin{align}
\label{eq:Limit1}
m\simeq 1-\frac{R_0(1-R_0 \overline{x_s^*})(x_s^*-\overline{x_s^*})}{(1-\overline{x_s^*})(1+R_0^2 \overline{x_s^*})},
\end{align}
with $\overline{x_s^*}\equiv -W_0(-R_0e^{-R_0})/R_0$ being the mean-field solution for $x_s^*$. Indeed, $m$ is close to $1$ in the vicinity of the maximum of the distribution,  $x_s^*\simeq \overline{x_s^*}$. Plugging Eq.(\ref{eq:Limit1}) into the action function Eq.(7), and approximating up to second order in $x_s^*-\overline{x_s^*}$, we find ${\cal S}(x_s^*)\simeq (1/2){\cal S}''(\;\overline{x_s^*}\;)(x_s^*-\overline{x_s^*}\;)^2$, with
\begin{align}
\label{eq:Limit2}
{\cal S}''(\;\overline{x_s^*}\;)=\frac{(R_0\overline{x_s^*}-1)^2}{(1-\overline{x_s^*}\;)\overline{x_s^*}(R_0^2\overline{x_s^*}+1)}.
\end{align}

\begin{figure}[t]
\center{\includegraphics[scale=0.18]{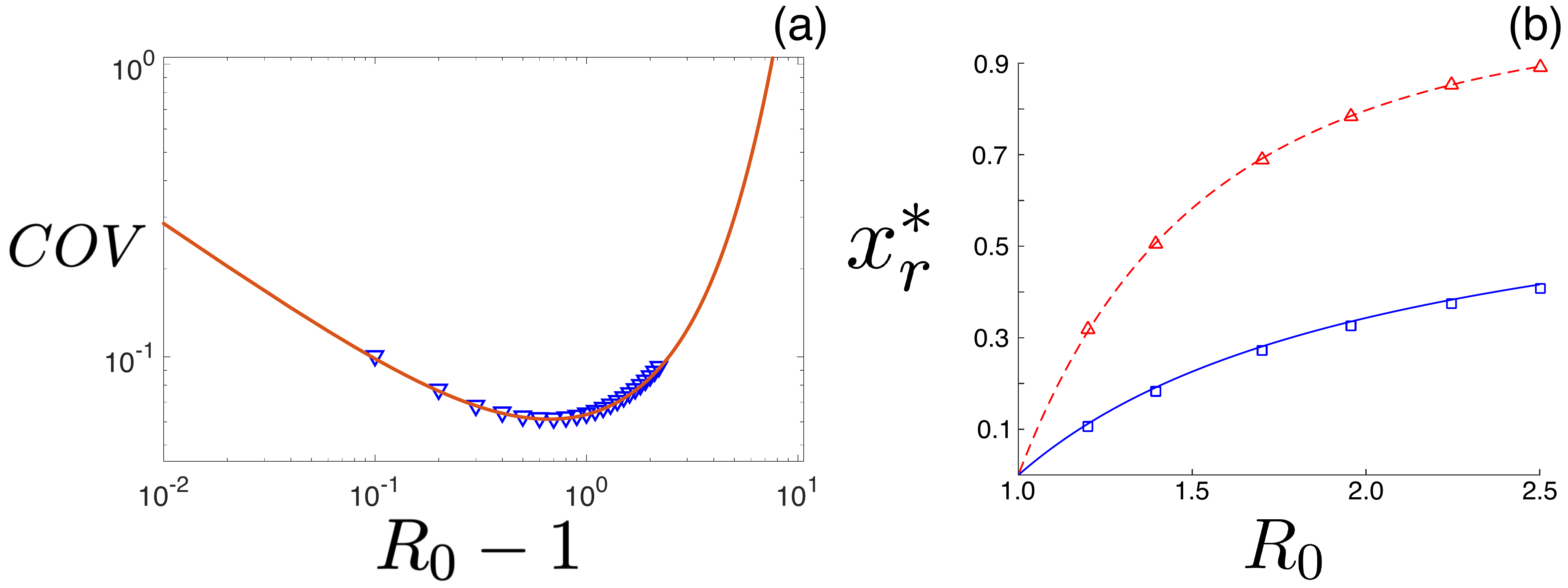}}
\caption{Shape of the outbreak distribution. (a) Ratio of the distribution standard deviation (around the mean field) to its mean vs. $R_0-1$ on a log-log scale. Symbols are solutions of Eq.(7) and $N=5000$, while the line is given by $\sigma/\;\overline{x_s^*}$ for the SIR model. (b) The least-likely (blue line) and most-likely (red line) outbreaks in the SEIR model versus $R_{0}\!=\!\beta/\gamma$, computed from Eq.~(7). Squares and triangles represent measured distribution minima and maxima from stochastic simulations. Population sizes were chosen so that $N\!{\cal S}(x_{r}^{*})\!=\!17$.  Other parameters are $\gamma\!=\!1$ and $\alpha\!=\!2$.}
\label{fig:EO2}
\end{figure}

Equation (\ref{eq:Limit2}) means that the distribution in the vicinity of the mean field outbreak size is a Gaussian with a width of $\sigma=(N {\cal S}''(\;\overline{x_s^*}\;))^{-1/2}$.
%(or, if one writes the distribution for the actual number of infected rather than the density, the width becomes $(N/{\cal S}''(x_0))^{1/2}$).
Notably, close to the bifurcation, $R_0-1\ll 1$,  $\overline{x_s^*}\simeq 1-2(R_0-1)$, and the width simplifies to $\sigma\simeq 2/\sqrt{N(R_0-1)}$, whereas for $R_0\gg 1$,  $\overline{x_s^*}\simeq e^{-R_0}$ and $\sigma\simeq N^{-1/2}e^{-R_0/2}$.
These calculations lead to a very interesting result: the coefficient of variation (COV), $COV=\sigma/\;\overline{x_s^*}$, receives a minimum at $R_0=1.66$. That is, the deviation from the mean-field outbreak size is minimized at $R_0\simeq 5/3$, whereas at $R_0\to 1$ or $R_0\gg 1$, the COV diverges, see Fig.~\ref{fig:EO2} (a).

Another unique aspect of the outbreak distribution is the least-likely small outbreak, $x_{r}^{\text{min}}$, which lies in between the mean field and the minimum outbreak $x_{r}^{*}\!=\!0$. The least-likely small outbreak satisfies $\partial {\cal S}/\partial x_{r}^{*}(x_{r}^{\text{min}})=0$ from the main text Eq.(7). For outbreaks smaller than the mean field, $x_{r}^{\text{min}}$ can be used to separate outbreaks into increasing or decreasing likelihoods. Usefully, we can track its dependence on parameters and compare to both Monte-Carlo simulations and the mean field.

An example is shown in Fig.\ref{fig:EO2} (b), where we plot $x_{r}^{\text{min}}$ and $\overline{x_r^*}$ versus $R_{0}$. The predicted value of $x_{r}^{\text{min}}$ (from solving Eq.(7)) is shown with a blue line, which can be compared directly with simulation results shown with blue squares. The latter were determined by first building histograms from $10^{11}$ stochastic simulations in the SEIR model, similar to Fig.(2) (a), and then fitting the smallest-probability region below the mean-field value with a quartic polynomial of $x_{r}^{*}$. The polynomial-fits were done to $log_{10}(P)$. After fitting, the local minimum was extracted for each plotted value of $R_{0}$. The least-likely small outbreaks can also be compared to the mean-field result, $\overline{x_r^*}$, shown in red. Similar to the blue series, lines are theory predictions and points represent the local maxima of the simulation-based histograms.

Note that we predicted that SEIR-model distributions are identical (on log scale) to SIR-model distributions, as described in the main text. This claim is tested in Fig.\ref{fig:EO2} (b), since simulations were performed under the former, while theory derived from the latter. As we can see, the two agree very well.

%\bibliography{ExtremeOutbreaks}


\end{document}